\documentclass[preprint,aps]{revtex4}
\usepackage{graphicx}
\usepackage{booktabs}
\usepackage[usenames, dvipsnames]{color}
\usepackage{multirow}
\usepackage{eurosym}
\usepackage{setspace}
\usepackage{rotating}
\usepackage[ pdftex, plainpages = false, pdfpagelabels, 
                 pdfpagelayout = useoutlines,
                 bookmarks,
                 bookmarksopen = true,
                 bookmarksnumbered = true,
                 breaklinks = true,
                 linktocpage=all,
                 pagebackref=false,
                 colorlinks = true,
                 linkcolor = BrickRed,
                 urlcolor  = blue,
                 citecolor = BrickRed,
                 anchorcolor = green,
                 hyperindex = true,
                 hyperfigures
                 ]{hyperref}

\begin{document}

\title{\href{http://www.necsi.edu/projects/yaneer/phenomenology}{The Limits of Phenomenology: From Behaviorism to Drug Testing and Engineering Design}}

\author{\href{http://necsi.edu/faculty/bar-yam.html}{Yaneer Bar-Yam}}
\affiliation{\href{http://www.necsi.edu}{New England Complex Systems Institute} \\ 
238 Main St. S319 Cambridge MA 02142, USA } 

\begin{abstract}
It is widely believed that theory is useful in physics because it describes simple systems and that strictly empirical phenomenological approaches are necessary for complex biological and social systems. Here we prove based upon an analysis of the information that can be obtained from experimental observations that theory is even more essential in the understanding of complex systems. Implications of this proof revise the general understanding of how we can understand complex systems including the behaviorist approach to human behavior, problems with testing engineered systems, and medical experimentation for evaluating treatments and the FDA approval of medications. Each of these approaches are inherently limited in their ability to characterize real world systems due to the large number of conditions that can affect their behavior. Models are necessary as they can help to characterize behavior without requiring observations for all possible conditions. The testing of models by empirical observations enhances the utility of those observations. For systems for which adequate models have not been developed, or are not practical, the limitations of empirical testing lead to uncertainty in our knowledge and risks in individual, organizational and social policy decisions. These risks should be recognized and inform our decisions.
\end{abstract}
\date{Aug. 1, 2013}

\maketitle

\section{Introduction}

The scientific method is often a subject of discussion \cite{kuhn,gauch} and pedagogy \cite{shipman,a4,a5}. Here we formalize and quantify the discussion of scientific method itself by applying information theory \cite{a6} to analyze the process of observation and description. The subject of our analysis will be a set of strictly defined empirical approaches. We include in this category classical behaviorism \cite{a7,a8,a9} and modern medical double-blind experiments \cite{a10,a11,a12}. These approaches adopt a specific relationship between observations, description and inference that are characterized by self-imposed limitations on the type of observations that are possible, and the relationship of experiment and theory. We show that when these limitations are imposed, the number of observations needed grows exponentially in the amount of information required to describe the conditions affecting the system. For any system that has more than a few possible conditions, the amount of information needed to describe it is not communicable in any reasonable time or writable on any reasonable medium. This result constrains the possible advances in knowledge. 

Our analysis is in the sprit of computational complexity theory in that it focuses on the scaling of the problem with its difficulty (complexity). We show that the scaling of a strictly empirical approach is exponential. Such exponential problems are impossible for all but the simplest problems, and alternative approaches must be identified. Our analysis may be considered to be an impossibility theorem, like the Halting problem or G\"odel's incompleteness theorems, in the context of epistemology---the study of the nature of knowledge.  However, our framing also provides insight about how progress can and even must be made. 

The analysis provides a guide to improving the process of medication approval, and the study of human psychology. Specifically, it shows that theoretical models are necessary for the characterization of complex systems. We also point to parallel insights for engineering, specifically the validation and testing of engineered systems---a significant practical concern. While the scaling limits of empirical approaches can be addressed by incorporating theory, the ability to empirically validate theory remains limited and where adequate theories have not been developed uncertainty translates into decision-making risks whose existence have far reaching implications.

A key contribution of our work is the unification of these various seemingly disparate issues within a single mathematical and conceptual framework through information analysis. Ashby has promoted the use of models as a practical matter pointing out the limitation of individuals in processing data \cite{a13}; it is likely that he  recognized that the same limitation applies to the processing of information by society and therefore to scientific analysis itself, which is the focus of our analysis.

Brief discussions of our analysis have been given previously \cite{DCS,YB2,complexeng}, as well as its implications for innovation and engineering of complex systems where failures that may be associated with the inability to test systems are common \cite{complexeng}. Recently this issue has been pointed to in the context of engineering of biological systems in regard to the risks associated with genetically modified organisms (GMOs) \cite{taleb}.

\section{Analysis}

In order to frame the discussion of scientific methodology we consider a set of individual agents (scientists) that are independently observing the behavior of a system. Progress is made when one scientist making observations communicates these observations to the other scientists. The subject of our analysis is the length of the message that provides a complete description of the system. (The number of scientists is not important to our analysis as even a single scientist must record findings, and the resulting record is the subject of our analysis.) We will analyze this message length for an assumed methodology of scientific inquiry. The scientists adopt a methodology that consists of determining the action (outcome, dependent variables) of the system in response to a mutually agreed upon set of conditions (see Fig. 1).  The set of conditions can include environmental conditions as well as internal conditions, whether controllable or uncontrollable---treatments, inputs, circumstances---constituting the combined set of values of all independent variables. This definition is consistent with the methodology of phenomenological approaches, including behaviorism and medical drug testing.

\begin{figure}[tb!]
\centering \includegraphics[width=12cm]{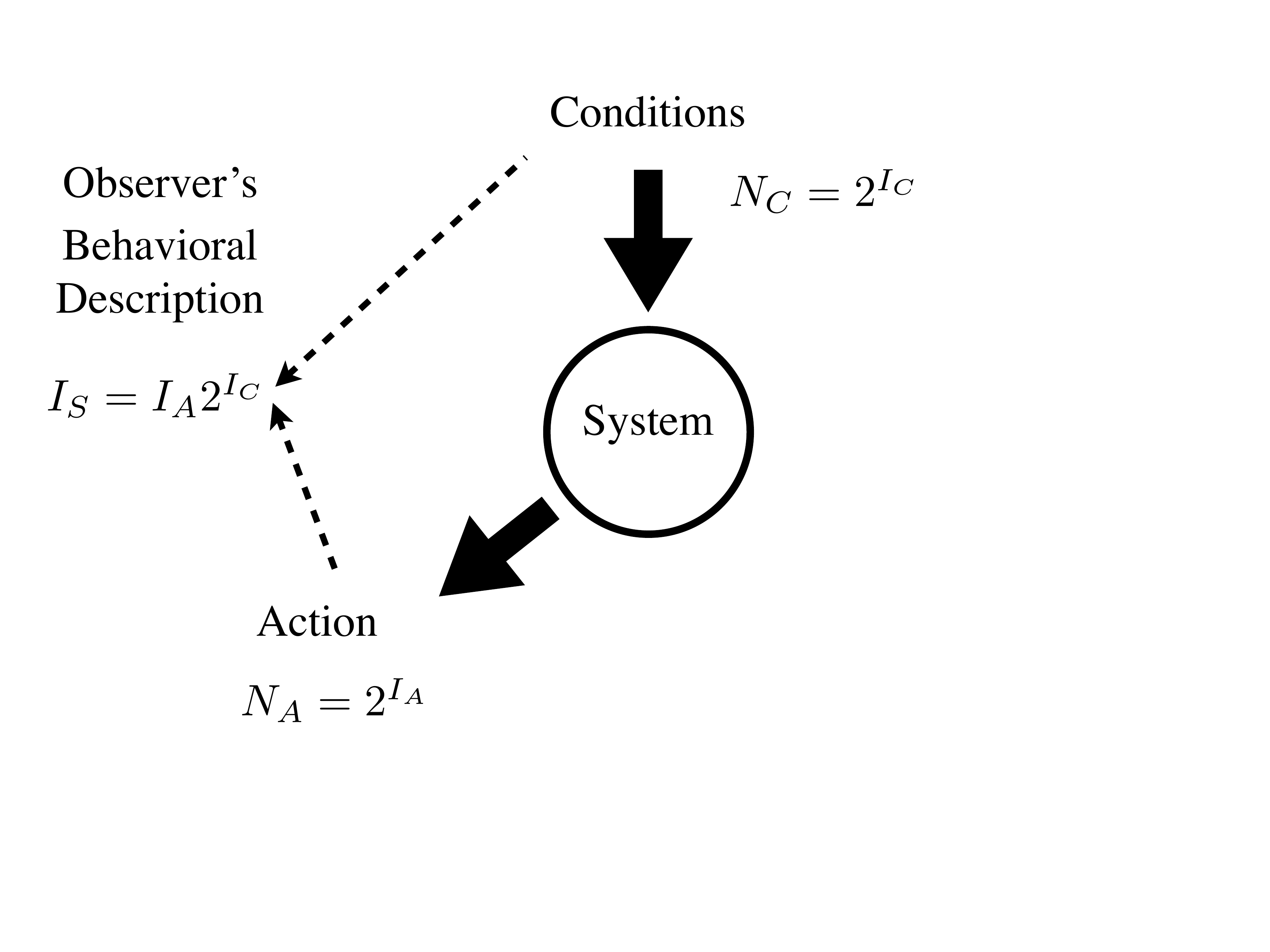}
\caption{\label{fig1} In describing the behavior of a system using a strictly empirical phenomenological approach, we observe a system as it responds (Action) to a set of external and internal conditions (Conditions). The number of possible responses is $N_A=2^{I_A}$ actions, where $I_A$ is the information needed to specify which action occured. The number of distinct conditions is $N_C=2^{I_C}$, where $I_C$ is the information necessary to specify the condition. Observers of the system can provide a behavioral description of the system by providing $I_S=I_A 2^{I_C}$ bits of information. Exponential growth of this information with $I_C$ implies such behavioral descriptions are impractical for all but small values of $I_C$, i.e. for a very limited set of conditions.}
\end{figure}

In an analysis based upon information theory [6] we identify the minimum amount of information necessary to communicate a full description of the system based upon the number of such descriptions that are possible. Thus if there are $N$ possible messages, the minimum amount of information that is needed is $I=\log_2(N)$ bits of information. This is required because each possible message must have a distinct representation and the number of possible messages with $I$ bits is $2^{I}$. Messages might wastefully use more than this number of bits but this is the minimum message length that enables a unique message and thus unambiguous communication.

In order to facilitate the communication between scientists engaged in studies of a particular system, we assume they jointly act in a specific way to minimize the amount of information that must be communicated. First, the scientists determine in advance the set of conditions that are to be characterized and assign a distinct numerical label for each of them. Thus, the only information that must be communicated by a scientist performing a single observation is the numerical label of the condition. The amount of information necessary in order to specify the condition being observed is only $I_C = \log_2(N_C)$. Communicating the result of the observation consists of identifying the action that resulted from that condition. Similar to the conditions, the set of actions of interest are identified in advance and are labeled numerically for identification. The amount of information necessary to specify the action is $I_A = \log_2(N_A)$. Communicating a single observation requires at least $I_C + I_A$ bits of information that specify in sequence which condition is being observed, and which action is found for that environment.

The primary object of our analysis is a complete description of the behavior of the system---a list of which action arises for each of the possible environmental conditions. The amount of information necessary to provide this description is given by:
\begin{equation}
I_S = 2^{I_C} I_A 
\end{equation}
This result follows from noting that for a complete description we need to specify an observation for each environmental condition of which there are $N_C=2^{I_C}$. This has to be multiplied by the amount of information necessary to communicate which action occurs, $I_A$. It might seem that each observation must be accompanied also by a specification of the environment. However, we can abbreviate a complete description  by recognizing that it is not necessary to communicate which environment is being observed, it is only necessary to order the actions observed in a pre-specified sequence. The action of the system in response to the first condition is listed first, the second second, and so on. The list has $N_C$ members, with each member of the list having $I_A$ bits of information. The total number of bits in the list is therefore given by Eq. (1). Note that the analysis can be extended for probabilistic outcomes by taking the description of the action to consist of a set of probabilities at a desired precision. Note also that we do not include the information needed to specify the correspondence of labels to the real world nature of the conditions or actions.

Eq. (1) can be stated as a rigorous ``behavioral complexity'' theorem determining the information in a description of a system's response: Given a set of conditions $N_C$ and a set of possible responses $N_A$, the amount of information necessary to describe the behavior of a system is given by Eq. (1), with $I_C$ and $I_A$ defined as before. This theorem is well known as the amount of information to characterize a boolean function that produces $I_A$ bits from $I_C$ input bits, the size of its truth table \cite{logic}. Our interest is in using this result to analyze the limitations of scientific characterization and methodology. The importance of Eq. (1) for scientific methodology resides primarily in the rapid growth of the amount of information with $I_C$. In the following, we illustrate this as a problem in specific well accepted methodologies.

\section{Discussion}

The mathematical discussion of behavioral complexity has direct implications for our understanding of scientific inquiry. Strictly empirical approaches that self-limit the nature of inquiry can be directly mapped onto the mathematical formulation of the theorem. 

Consider behaviorism. Behaviorism, as defined in the early 20th century is the description of behavior without reference to the internal functioning of the brain. Instead, as epitomized by Pavlov's research on dogs [7] and Skinner's box [8], the organism (human or animal) is presented with a specified set of environmental conditions. The response to each of these environmental conditions is measured. This set of responses then is considered to characterize the behavior of the system. Of key importance to our application of the theorem to this scientific approach is its self-imposed exclusion of considering the internal functioning of the system. The traditional experiments that were used to establish the discussion of behaviorism were limited to very few environmental conditions. For example, for Pavlov's experiments measuring the conditioned reflex of dog salivation to bell ringing, resulting from prior bell ringing when eating food, an enumeration of conditions is: 
\begin{itemize}
\item no prior bell ringing with food, no bell ringing, no food present,
\item no prior bell ringing with food, no bell ringing, food present, 
\item no prior bell ringing with food, bell ringing, no food present,
\item no prior bell ringing with food, bell ringing, food present,
\item prior bell ringing with food, no bell ringing, no food present,
\item prior bell ringing with food, no bell ringing, food present, 
\item prior bell ringing with food, bell ringing, no food present, and
\item prior bell ringing with food, bell ringing, food present.
\end{itemize}
\noindent The possible responses were 
\begin{itemize}
\item salivating, and
\item not salivating. 
\end{itemize}
\noindent Similarly only a few conditions were present in Skinner's box---indeed the whole point of having a ``box'' is to restrict severely the set of possible environmental conditions. Difficulties arise when we consider the prospect of extending this approach as a method for scientific study of all of human (or animal) behavior. In order to do this we must consider the informational demands on describing human response to all possible conditions that affect human behavior. While we do not have a complete characterization of what affects human behavior, we can consider the example of a human being reading and responding to written statements. A single sentence of a multiple-sentence written statement contains almost 200 letters and an information per letter of approximately 1 bit \cite{a6,a14}. We conclude that $I_C$, the amount of information in a written statement that a person responds to, can be greater than $200$ bits and the result is that the amount of information in the description given by Eq. (1) is greater than $2^{200}=10^{60}$ which is roughly the number of atoms in 1,000 suns \cite{a6,a14}. It is therefore reasonable to state that a project using phenomenological approaches to characterize human response to written communications would be impractical. 

The problem has to do with the scaling of the methodology as the demands increase. For a few cases, as in Pavlov's experiment, the method can be effective. However, as the desire arises to provide a better description that takes into consideration more possible factors, allowing for more diverse conditions, the amount of effort increases. This increase scales exponentially so that as the effort to understand system response to the environment increases, the methodology becomes rapidly impractical. 

Consider medical double blind experiments, the standard for determining treatment protocols for disease conditions and FDA approval of medications [10]. This paradigm directly maps onto our framework. In this methodology, a population that has a specific condition is identified. The population is separated into subsets, each of which receives a certain treatment or no treatment, and the outcomes are measured at a specified time later. In the simplest case there are two groups, a treated and untreated group, and the measure has only two possibilities, cured or not, which are definite outcomes. The information needed to specify the environment consists of a single bit, $\{0,1\}=\{untreated, treated\}$, and the observation consists of a single bit, $\{0,1\}=\{cured, not cured\}$. The amount of information necessary to specify the result in this case from Eq. (1) is two bits of information, i.e. which of the outcomes occurs for each of the two conditions. Such experiments serve as a core of medical knowledge. Our concern here is how this approach generalizes to more complex cases. In more complex cases there are both a larger set of conditions and a larger set of outcomes that are measured. Among the factors that increase the number of conditions are different conditions associated with the diseases (i.e. severity), but also additional characteristics such as gender, age cohort, co-occurring conditions such as pregnancy, other diseases (such as heart conditions when treating for depression \cite{a15}), drug interactions, etc.  Among the larger set of outcomes are, for example, probabilities of survival, relief of certain specified symptoms, and probabilities of side effects. Adding cases to the outcomes is not as problematic as adding more possibilities to the set of conditions. As can be seen from Eq. (1) we must have a distinct observation for each of the conditions that is to be observed, i.e. there is need for one independent population for each of the conditions to be tested, a population which is then also large enough to represent the possible outcomes at the level of precision desired. 

Such an approach cannot accommodate more than a few intrinsic conditions or the large number of conditions for possible drug interactions. The limitations of this approach from a mathematical perspective are manifest in difficulties that have been experienced over the past two decades in the conflict between adequate testing of medications in the face of cautions about side effects and low probability adverse outcomes, and the need to bring new medications to use \cite{a15,a16,a17}. 

Consider how the same principle appears in engineering, which in principle is an inversely related process. The idea of engineering is to design and build a system that performs a particular pre-specified set of responses. In a standard sequence of events, the set of responses is decided upon, a design is constructed, it is implemented in hardware or software, and the result is then implemented for use. However there is a step before use that consists of testing. What is the framework for the testing? When the system is tested we have an intended functional description, and we have the system. What remains is to compare them. The difficulty is that we need to perform observations of the response in order to test them. Thus testing has a similar constraint to strict empirical scientific methods---we are not allowed to look inside the system in order to evaluate its ability to perform the behavior; we want to see it perform the behavior. How many bits of information must we collect in order to test it? This is specified by Eq. (1) --- for each condition we must evaluate its behavior and see if that behavior matches the desired one. Our theorem predicts that when there are sufficiently many bits of information necessary to describe the conditions, the testing process becomes impractical.

We can evaluate this for the case of computer chip design. A computer chip is specified in terms of an algorithm that relates input pins to output pins. The pins are subject to voltages that can be considered as binary, either ON or OFF.  The number of input pins represents the amount of environmental information, and the number of output pins represents the amount of behavioral response of the system identified as the computer chip. The problem of testing is then to match a given description of the input-output mapping of the computer chip to the actual performance of that chip. According to our fundamental principle, this process requires $I_A 2^{I_C}$ tests, and even if some assumptions are made, it surely should not be possible to adequately test a chip that has more than 200 bits of input information due to the huge number of potential tests. 

This prediction can be compared to the experience with actual chip design. The problems with ``bugs'' in chips after extensive testing originally became a prominent issue during the 1990s when multiple Intel processor chips appeared with errors  \cite{a18,a19,a20}. The number of pins in the Intel Socket 4 for pentium processors is 273 (Socket 3 is 237 pins). That an error occurred despite the recognized need for testing demonstrates an inherent difficulty. This problem was overcome by design for testability---making multiple modules each of which has fewer pins and can be tested and the joining of multiple components is itself testable \cite{testing}. While computer chips can be readily modularized for testability, this solution is not readily applied to all systems.  The failure rate of complex engineered systems has been very high in recent years. Methods to address this problem apply evolutionary trial and error testing \cite{complexeng,complexeng1,complexeng2}. The irony of modern day approaches to engineering of biological systems, including GMOs, is that it is regressive in using the more traditional engineering strategy that does not recognize the associated risks \cite{taleb}.

We can apply this discussion to the role of databases in the study of complex biological or social systems today. A proliferation of phenomenological databases is occurring with hundreds of gigabytes or terabytes of information about measurements about systems, such as high throughput data in biology or ``big data'' in society. We can see that our analysis formalizes the question: How much data must be obtained to provide the desired information about a system? Our analysis shows that for a complex system, the amount of data is exponentially large in the amount of information describing the conditions, and linear in the amount of information describing actions. We can use this lower bound estimate to evaluate whether a strictly data oriented approach is practical for a given system.

In each of these circumstances, we find that the approach can be summarized in the following way. There is a system, $S$, which is being observed. There are a set of conditions that are being applied to the system, and a set of possible outcomes of interest. To characterize this situation in general, we do not need to identify the specific nature of each of the conditions---what is important is the number of these environmental conditions $N_C$ and the number of the possible outcomes or actions of the system $N_A$. The key concept is a calculation from these of the amount of information necessary to specify the behavior of the system. 

What is the alternative to the strict phenomenological approach in science? How can we reduce the amount of information needed? The only way to do so is to characterize the response of the system to the environments through a concise representation of the system's behavior, i.e. a model. In order to represent the behavior of a system in response to a large set of possible conditions, it is necessary to have a representation of the system that identifies the response to cases that have not been observed. A model enables inference to determine the behavior of a system across a range of conditions. 

Given the large number of possible observations, observed cases are only a sparse sampling of the cases that are possible. Any individual observation can only serve as a test of the model rather than in itself be a representation of the system's response. The more complex the system is, i.e. the more it can respond to different possible environmental conditions, the more the interplay of model and testing becomes critical to the scientific approach. Theoretical approaches are often used in biological and social systems as well as in physical ones. Our analysis identifies the need for such approaches in contrast to strict phenomenological ones.

The analysis of our theorem implies that a phenomenological approach is ineffective because of the large number of observations needed. Models are necessary. However, this realization does not eliminate the limitations associated with testing. The mathematical analysis implies that our knowledge is inherently limited by the ability to evaluate only a few of the large number of conditions of response of a system. That effective empirical strategies are ultimately bound to models implies that both are subject to the same limitations, and therefore certainty is always bounded. 
We cannot use strictly empirical approaches to eliminate the uncertainty associated with modeling extrapolations from existing observations to those that we have not done. Modeling is necessary, certainty is not possible. Limitations of certainty should enter into discussions of risks being taken and estimations of risk have implications for effective decision making in individual, organizational and social policy contexts. 

An example of a commonly used type of simple model is  that the response of the system to its environments is independent of some aspect of the environment. For example, assume that the conditions include exposure to a number of different factors. We could assume that the action of the system only depends on one of them. This would then simplify our understanding of the response of the system. This type of model is what is commonly used for strictly empirical studies. Indeed, we can now state that strict empiricism is not and has never been ``model free.'' Instead it uses a very limited kind of theory in which the role of certain factors as the only causes of specific outcomes is assumed rather than tested.

What this means in practical terms is that any laboratory experiment in biology must use and indeed is using a controlled environment in which to perform the experiment. This controlling itself limits the study as a method for understanding the behavior of the system. Similarly, strictly empirical medical or social studies are limited by the applicability of their choice of restrictions on the conditions that are being evaluated. 

Another type of simple model commonly used assumes that intermediate values of conditions have intermediate outcomes according to some metric. Linear, or more generally, smooth interpolation is a model of system behavior that may or may not be well-justified. By recognizing explicitly when such simple models are being used, one can question them and replace them with improved ones as new information is obtained. 

The approach of using a more concise model to represent the behavior of a system may be thought of as a kind of data compression of the information about the system. However, there is an important distinction between models and compression. If we don't actually collect all of the observations about the system (because it is not practical) then it is a model/theory.  If we collect the data but compress it prior to communication then it is a strictly empirical strategy. Compression reduces the ultimate communication, but doesn't reduce the amount of information gathered. 

In what way is a model different from a falsifiable hypothesis? A hypothesis is often a conceptual understanding that provides insight into, i.e. predict, one or more  observations. In order to fulfill the criteria we have established here, a model must provide predictions of the results of a large number of observations. In some cases models are used to describe observations after they have been performed, and are not considered to be applicable beyond them. Such models do not satisfy our criterion. It is the ability to identify the results of observations that have not yet been performed that is necessary.

\section{Conclusions}

In summary, behaviorism and related phenomenological approaches that do not use inference to determine behavior from more limited information cannot characterize complex systems. We can define complex systems for these purposes as systems that respond to many bits of information as described by Eq. (1). In order to describe the behaviors of such a system, a validated model that generates behaviors from simpler characterizations must be used. 

Thus, it is precisely for highly complex biological and social systems that theoretical modeling is essential to the scientific process. Existing approaches to biological and social sciences include theoretical and empirical approaches with various levels of integration between them. Our analysis discounting the effectiveness of empirical approaches applies to strict empirical framings of research that command much of the attention of scientists despite the limitations that we are describing here. 

The need for an alternative approach in medicine is increasingly manifest as personalized medications based upon genetic testing are in development \cite{personalmedicine,huber}. Since there is no population of individuals on which such treatments can be tested, the inadequacy of traditional methods is apparent. Nevertheless, current framings of these approaches implicitly use models whose assumptions are not being acknowledged or validated.

We can strengthen our discussion further, based upon an additional statement that given an exponentially large set of possible environmental conditions, the chance that any particular condition will recur more than once is vanishingly small. While this may not be the case for particular artificial constructs such as computer chips, real world biological and social systems have levels of detail to their set of conditions that are not measurable. Even if they are measurable, the fine granularity implies coincidence of two circumstances that are exactly the same is negligible. With this additional assumption we can state that:

No empirical observation is ever useful as a direct measure of a future observation. It is only through generalization motivated by some form of model/theory that we can use past information to address future circumstances. 

Under these conditions theory is necessary not just for a complete description of the behavior of a system, but also for the ability to identify the outcome for any new observation, given a set of prior empirical observations. Why therefore, are strictly empirical approaches considered to be effective---because they inherently adopt, without testing, simplifying assumptions. Only a few conditions are identified, within each of these conditions changes in other factors are considered not to be relevant to the outcomes being considered, and therefore conditions are replicable.
It is interesting to point out in this context that medical journals do not accept models based upon the assumption that they are weaker forms of scientific knowledge than ``actual observations.'' However, it is the assumptions of replicability, independence and transferability that are ultimately the weaker assumptions. The current dominant approach to medical knowledge is limited due to its methodology in adopting the limitations of empiricism as a weak form of theory. It is only through the assumption that there are only a few important conditions that the use of past observations can be considered to be applicable to future ones. 

Consider the statement that ``all theories/models are false \cite{a21}.'' While this is a reasonable statement, we see from our discussion that it misses an essential point. Theories are essential; strictly empirical approaches are  largely untested, poorly justified, theories. So it should be understood that not just all theories that are false---all expectations about future events are in the same sense unreliable (i.e. false), since only in idealized simple systems can conditions be replicated. For complex systems all empirical inferences are just as false, indeed they are actually more limited than theories by their assumptions of replicability of conditions, independence of different causal factors, and transfer to different conditions of prior observations. Thus we cannot rely upon their validity. The approach that is more effective is to develop better models that can identify outcomes of a wide range of conditions, and empirically test them as much as possible. 

The importance of theory does not mean that there is a diminished importance of data. Indeed, for complex systems it is not possible to have enough data to fully describe the system. This means that we have to use the data we do have as effectively as possible. An important way to understand the role of theory is that it makes data more useful. By using data to test the theory, it can be generalized to conditions that have not yet been observed.  The advent of ``big data'' is critical for addressing complex systems, but without recognizing the sparseness of that data in an exponentially rich possible data set we are limited in the progress we can make.  Observations should be designed (when possible) to optimally sample the space of possibilities so that the best theoretical inferences can be obtained about all conditions not just the ones that have been seen. Controlled experiments may be particularly useful because they enable more optimal testing of theories. Data collected without controls provides an alternative and increasingly important source of information relevant to testing of theories, but where optimization may not be possible.

Given the severe limitations that we have identified of strictly empirical approaches we should ask why have we been able to make as much progress as we have until now? The human ability to create mental models provides an explanation. This model-making capability and the resulting ``intuition'' has been a large part of our advances rather than the more limited empirical strategies as well as the limited mathematical apparatus of calculus and statistics. While traditional mathematical theories have been confined to the assumptions of calculus and statistics, human model-making is not, and neither are new theoretical methods \cite{DCS}.

Recognizing that real world conditions cannot be replicated, our ability to identify important factors that influence outcomes is essential. Ultimately, the possibility of knowledge itself must rely upon the ability to differentiate between different pieces of information based on their importance. Observations must focus on those pieces of information and not on the rest. Without such an ability, the listing of always irrelevant facts is unavoidable. Thus it is evident that an essential role of theory must be to identify which pieces of information are important. We must therefore ask whether we have tools to do the former, a discussion of which is the subject of a separate article \cite{a22}. It is key to understand how and for what questions it is possible to overcome the limitations we have identified. Prior to this discussion, it is reasonable to suggest that in the future we should both use the best theoretical tools and make the most use of our ``intuition'' (i.e. mental models) as model-making systems. Practical approaches to medicine, management and policy require a better framing of how we can effectively understand biological and social systems. Recognizing that empirical approaches do not extend well to complex systems, and that theory and experiment must work hand in hand is an important step in the right direction. 

\section{Appendix}

For reference, we include separately here the central theorem and two corollaries.

\subsection*{Fundamental theorem of complex systems: Behavioral complexity}

Given a system whose behavior we want to specify, for which the conditions as input variables have a complexity of $I_C$, and the actions of the system have a complexity of $I_A$, then the complexity of specification of the behavior of the system is:
\begin{equation}
I_S=2^{I_C} I_A 
\end{equation}
Where complexity is defined as the logarithm (base 2) of the number of possibilities or, equivalently, the length of a description in bits. The proof follows from recognizing that a complete specification of the behavior is given by a table whose rows are the actions ($I_A$ bits) for each possible input, of which there are $2^{I_C}$. Since no restriction has been assumed on the actions, all actions are possible and this is the minimal length description of the function. Note that this theorem applies to the complexity of description as defined by the observer, so that each of the quantities can be defined by the desires of the observer for descriptive accuracy.

\subsection*{Corollary: A complex system's responses are not independent}

According to physics, the amount of information necessary to describe completely a human being is no more than the entropy of a corresponding equilibrium system, with the same matter, volume and temperature, measured in bits. We can compare the entropy of a human being in bits with Eq (1) for human response to the text of a sentence. The entropy can be  estimated to be $10^{31}$ [14], which is much smaller than $10^{60}$. Thus, we can conclude that the information necessary to describe the quantum limited amount of information about a system would be more concise a description than a purely empirical behavioral one. Since, according to physics, the entropy is the maximum available information about a system, the responses of a system must be describable using a more concise description than Eq. (1). Equivalently, we can say that the responses of the system are not independent of each other.  This corollary does not provide us with the amount of information needed for a model, only that there exists models that require much less information than a strictly empirical approach. 

\subsection*{Corollary: Strictly empirical approaches (Behaviorism and Medical testing) are not practical for complex systems}

Since behaviorism and medical testing approaches self-impose the conditions of the theorem, the information needed to pursue them grows exponentially, and they are therefore impractical for all but the most limited sets of conditions. 

\subsection*{Technical note on the dynamics of progressive analysis}

In our analysis there is an {\em a-priori} assumed set of conditions and actions that are being studied. This framing is convenient for analysis. In the real world identifying the conditions and actions is part of the scientific process. While this creates a cyclical situation that is problematic for logic, it is overcome, like many other paradoxes in logic, by allowing for a dynamical process. Specifically we assume a progression of stages indexed by $\tau$ such that $I_C(\tau)$ and $I_A(\tau)$ are the $\tau$th scientific stages of inquiry. The first stage can be done arbitrarily.

\section{Acknowledgements}
Thanks to Casey Friedman, Maya Bialik and Roozbeh Daneshvar for helpful comments and Nassim Taleb for the motivation to complete this longstanding manuscript.


\begin{thebibliography}{200}

\bibitem{kuhn} TS Kuhn, The structure of scientific revolutions, 3rd edition (University of Chicago Press, Chicago, 1996).

\bibitem{gauch} HG Gauch Jr, Scientific method in practice (Cambridge University Press, Cambridge UK, 2003).

\bibitem{shipman} JT Shipman, JD Wilson, AW Todd, An introduction to physical science. (Houghton Mifflin, Boston, 2009). Pg. 3-4.

\bibitem{a4} SS Zumdahl, SA Zumdahl, Chemistry (Houghton Mifflin, Boston, 2007). Pg. 5-7.

\bibitem{a5} L Berg, Introductory botany: Plants, people, and the environment (Thomson Brooks/Cole, Belmont CA, 2008). Pg. 16-19.

\bibitem{a6} CE Shannon, A mathematical theory of communication. \emph{Bell System Technical Journal} \textbf{27}, 379-423, 623-656 (1948).

\bibitem{a7} IP Pavlov, Conditioned reflexes: An investigation of the physiological activity of the cerebral cortex (Oxford University Press, London, 1927).

\bibitem{a8} BF Skinner, About behaviorism (Random House Inc, New York, 1976).

\bibitem{a9} WM Baum, Understanding behaviorism: Behavior, culture, and evolution (Blackwell Publishing, Malden MA, 2005).

\bibitem{a10} Food and Drug Administration, Guidance for industry: E 10 Choice of control group and related issues in clinical trials (May 2001). \url{http://www.fda.gov/downloads/regulatoryinformation/guidances/ucm125912.pdf}

\bibitem{a11} National Institutes of Health, Standards for clinical research within the NIH Intramural Research Program (2009). \url{http://www.cc.nih.gov/ccc/clinicalresearch/standards1.html\#standards}

\bibitem{a12} KF Schulz, Altman DG, D Moher for the CONSORT Group, CONSORT 2010 Statement: Updated guidelines for reporting parallel group randomised trials, \emph{BMJ} \textbf{340}, c332 (2010).

\bibitem{a13} WR Ashby, Analysis of the system to be modeled, in R. M. Stogdill (Ed), The process of model-building in the behavioral sciences (W. W. Norton, New York, 1972).

\bibitem{DCS} Y Bar-Yam, Dynamics of complex systems (Westview Press, Boulder CO, 1997), Chapter 8.

\bibitem{YB2} Y Bar-Yam, Unifying principles in complex systems, in MC Roco, WS Bainbridge (Eds), Converging technology (NBIC) for improving human performance (Kluwer, New York, 2003). \url{http://necsi.edu/projects/yaneer/complexsystems.pdf}

\bibitem{complexeng} Y Bar-Yam, Large scale engineering and evolutionary change: Useful concepts for implementation of FORCEnet, \emph{Report to Chief of Naval Operations Strategic Studies Group} (2002).

\bibitem{taleb} NN Taleb, The Artangle Longplayer letters: To Stewart Brand ( Apr. 30, 2013). \url{http://longplayer.org/what/whatelse/letters.php} 

\bibitem{logic} HB Enderton, A mathematical introduction to logic. (Academic Press, New York, 1972).

\bibitem{a14} Y Bar-Yam, Dynamics of complex systems (Westview Press, Boulder CO, 1997), Chapter 1.8.

\bibitem{a15} Food and Drug Administration, FDA drug safety communication: Revised recommendations for Celexa (citalopram hydrobromide) related to a potential risk of abnormal heart rhythms with high doses. \emph{FDA Advisory} (2012). \url{http://www.fda.gov/drugs/drugsafety/ucm297391.htm}

\bibitem{a16} TM Burton, FDA urged to speed approval of drugs. \emph{Wall Street Journal} (Sept. 25, 2012) \url{http://online.wsj.com/article/SB10000872396390444083304578018790623838634.html}

\bibitem{a17} Food and Drug Administration, Fast track, accelerated approval and priority review (2012). \url{http://www.fda.gov/ForConsumers/ByAudience/ForPatientAdvocates/SpeedingAccesstoImportantNewTherapies/ucm128291.htm}

\bibitem{a18} J Markoff, Chip error continuing to dog officials at Intel. \emph{New York Times} (Dec. 6, 1994). \url{http://www.nytimes.com/1994/12/06/business/chip-error-continuing-to-dog-officials-at-intel.html}

\bibitem{a19} Mac Observer, Intel core chip bugs may affect Macs. \emph{Mac Observer} (Jan. 23, 2006). \url{http://www.macobserver.com/article/2006/01/23.6.shtml}

\bibitem{a20} Intel 8080, Wikipedia (accessed Nov. 14, 2012). \url{http://en.wikipedia.org/wiki/Intel_8080}

\bibitem{testing} EDA for IC system design, verification, and testing, Volume 1.

\bibitem{complexeng1} D Braha, A Minai, Y Bar-Yam (Eds), Complex engineered systems (Springer, Berlin, 2006).

\bibitem{complexeng2} DO Norman, ML Kuras, Engineering complex systems (MITRE, Bedford MA, 2004). 

\bibitem{personalmedicine} LJ van't Veer, R Bernards, Enabling personalized cancer medicine through analysis of gene-expression patterns, \emph{Nature} \textbf{452}, 564-570 (2008).

\bibitem{huber} P Huber, The digital future of molecular medicine: Rethinking FDA regulation, Manhattan Institute for Policy Research (May 2013).  \url{http://en.wikipedia.org/wiki/Intel_8080}

\bibitem{a21} J Sterman, All models are wrong: Reflections on becoming a systems scientist. \emph{System Dynamics Review} 18, 501-531 (2002).

\bibitem{a22} Y Bar-Yam, Beyond data: Identifying the important information for meeting real world challenges (in preparation).
 
\end{thebibliography}
\end{document}